\documentclass[12pt]{iopart}
\usepackage{geometry}                
\usepackage{graphicx}
\usepackage{amssymb}
\usepackage{epstopdf}

\begin{document}

\title[Casimir energy in a spherical surface within surface impedance approach]{Casimir energy in a spherical surface within surface impedance approach:  the Drude model}
\author{Luigi Rosa$^{1,2,\dagger}$, Lucia Trozzo$^{*}$}

\address{$^1$Dipartimento di Scienze Fisiche, Universit\`a di Napoli Federico II,\\
 Complesso Universitario di Monte Sant'Angelo, \\ Via Cintia Edificio 6, 80126 Napoli, Italy \\
$^2$ INFN, Sezione di Napoli, Complesso Universitario di Monte Sant'Angelo, \\
Via Cintia Edificio 6, 80126 Napoli, Italy}

\ead{$\dagger$ rosa@na.infn.it,\\
$*$lucia.trozzo2603@gmail.com
}

\date{}                                           

\begin{abstract}
The Casimir Energy of a spherical surface characterized by its surface impedance is calculated.
The material properties of the boundary are described by means of the Drude model, so that a generalization of  previous results is obtained.
The limits of the proposed approach are analyzed and a solution is suggested.
The possibility of modulating the sign of the Casimir force from positive (repulsion) to negative (attraction) is studied.
\end{abstract}


\maketitle

\section{Introduction}

Casimir force is one of the few macroscopic manifestations of quantum mechanics. Indeed the (generally attractive) force between two parallel conducting plates in vacuum is directly connected to the vacuum fluctuations of the electro-magnetic field within the plates \cite{Cas48}. Because the Casimir force becomes dominant at the nanometer scale it could constitute a strong limitation in  the production of nanodevices \cite{Buks01}. For this reasons, during last years people tried to understand under what conditions it is possible to change the sign of the force from attractive to repulsive \cite{Kenneth:2002ij}.

It was Boyer \cite{boyer74} in 1974 the first one to obtain a repulsive Casimir force between two plates, one perfectly conducting and the other infinitely permeable (see also \cite{Lambrecht:2008cy,Pirozhenko:2008tr,Rosa:2010wp,Alves:2000}).
In  \cite{Kenneth:2002ij} the authors showed that, under suitable conditions, the transition between attractive and repulsive regime only depends  on the Surface Impedance (SI) of the material constituting the boundary. Since then a lot of effort in that direction has been done, see \cite{Cocoletzi:1989lr,Geyer:2003zz,Esquivel03,Rosa:2012hj} and references there in.
As far as we know the first ones to use the SI to compute Casimir energy were Mostepanenko and Trunov in \cite{Moste85}.

In this paper we follow a twofold line of reasoning, from one side we want to extend a previous analysis \cite{Rosa:2012hj} (to which we refer for details) to a more general case.  At the same time, we would like to understand the formal limitation of that approach (if any), see conclusions in \cite{Rosa:2012hj}. In this way, studying  the structure of the divergencies appearing, we succeed in fixing some computation errors present in \cite{Rosa:2012hj}, and we clarify what are the  difficulties encountered.
 
The paper is organized as follows: in section 2 the general setup of the calculation is introduced, in section 3 it is applied to the case under consideration, and  in section 4 we give our  conclusions. In the appendix we collect some relevant  formulae. 
 
\section{ Casimir energy and the zeta function regularization}
 
The Casimir energy is the vacuum expectation value of the Hamiltonian operator
 on the ground state  \cite{ Bordag:2009zzd,Plunien:1986ca}:
  \begin{equation}
E_{Cas}=\left\langle 0\left|H\right|0\right\rangle =\sum_{J}\frac{1}{2}E_{J}\label{eq:defencas}
\end{equation}
where  $E_{J}$ are the energy eigenvalues labeled by some index $J$.
In general the sum is divergent and a regularization  is necessary. In the following we will use the zeta-function regularization \cite{Bordag:2009zzd}. In this approach a convenient exponent $s$ is introduced in the sum \eref{eq:defencas} so to make it convergent. The final result is recovered by a limiting procedure:
\begin{equation}
E_{Cas}=\lim_{s\rightarrow-1/2}\frac{\mu^{2s+1}}{2}\sum E_{J}^{(-2s)}=:\lim_{s\rightarrow-1/2}\mu^{2s+1}\zeta_{H}(s)\label{eq:encasreg}
\end{equation}
where  $\zeta_{H}(s)$ is the Riemann zeta function relative to the  operator  $H$ and $\mu$ is a dimensional parameter introduced so to have $E_{J}$ adimensional.
It will disappear on removing the regularization in the limit   $s\rightarrow-1/2$.
 
The main problem is that no explicit expression for the eigenvalues exists.
One way of overcoming this difficulty is by  means of  the Cauchy argument principle \cite{Ahlfors79}: 

{\em
 If $\Delta$ is analytic in a region $\Omega$, except for poles $x$ and zeros $y$, and $g(z)$ is analytic in 
$\Omega$ then}
\begin{equation}
\sum_{l}g(x_{l})-\sum_{m}g(y_{m})=\frac{1}{2\pi i}\oint_{\gamma}g(k)\log(\Delta(k))\label{eq:princarg}
\end{equation}
{\em where   $\gamma$ is a closed contour in $\Omega$ that does not pass through any of the zeros and poles, and contains all the zeros  $x_{l}$ and poles $y_{l}$ of the function  $\Delta$.}

Thus, choosing the function  $\Delta(k)$ such that its zeros coincide with the eigenvalues of our problem and has no poles inside  $\gamma$ we have
\begin{equation}
\zeta_{H}=\sum_{J}\frac{1}{2\pi i}\oint_{\gamma}g(k)\log(\Delta_{J}(k))\label{eq:zeta1}.
\end{equation}
For this reason $\Delta$ is called the generating function. The sum over  $J$
takes into account possible degeneracy of the eigenvalues.

In the following we will use the SI to obtain the generating function.
The SI  of a surface material $(\Sigma)$ is defined through the equation
\cite{Bordag:2009zzd,Jackson98}
\begin{equation}
\mathbf{E}_{t}\mid_{\Sigma}=Z(\mathbf{H}\times n)\mid_{\Sigma}\label{eq:zimp}
\end{equation}
where  $n$ is the outward normal to the surface. It relates the tangential components of the fields outside the material whose properties are encoded in $Z$ \cite{Stratton:1941kx}. The big advantage is that, in all the cases in which the dielectric properties of the material $\epsilon,~ \mu$ are known, a direct relation exists between them and $Z$:
\begin{equation*}
Z=\sqrt{\frac{\mu}{\varepsilon}}\label{eq:zimp2}
\end{equation*}
otherwise equation (\ref{eq:zimp}) can be viewed as a functional definition for the SI (see \cite{Esquivel03} and references there in).

\section{The Casimir Energy for a sphere}

In this section, we will concentrate on the case of the electromagnetic field in a sphere of radius  $a$. By means of \eref{eq:zimp} we will obtain the  generating function $\Delta_{J}(k)$ so to compute the $\zeta_{H}$. 

 The electric and magnetic field within the sphere can be written as \cite{Bordag:2009zzd,Jackson98,Mie908,Debye909,Bowkamp54}
\begin{eqnarray}
\mathbf{E} &=& \sum_{l=1}^{\infty}\frac{i}{ka}A^{TE}[i\mathbf{\hat{n}}j_{\nu}(kr)Y_{l,m}(\theta,\varphi)+(krj_{\nu}(kr))'\hat{n}\times\mathbf{X}_{l,m}(\theta,\varphi)]+\nonumber \\
& &A^{TM}j_{\nu}(kr)\mathbf{X}_{l,m}(\theta,\varphi)\\
\mathbf{H} &=& \sum_{l=1}^{\infty}\frac{k}{\omega\mu}\{A^{TE}j_{\nu}(kr)\mathbf{X}_{l,m}(\theta,\varphi)
-i\frac{A^{TM}}{rk}[i\mathbf{\hat{n}}j_{\nu}(kr)Y_{l,m}(\theta,\varphi)+ \nonumber \\
& &(krj_{\nu}(kr))'\hat{n}\times\mathbf{X}_{l,m}(\theta,\varphi)] \label{eq:max6}
\end{eqnarray}
where  $Y_{l,m}$ and  $\mathbf{X}_{l,m}$ are the scalar and the vectorial spherical harmonics respectively \cite{Jackson98}, $k=\sqrt{\epsilon\mu}\omega$ and $j_{\nu}(x)=\sqrt{\frac{\pi}{2x}}J_{\nu}(x)$ \cite{Abramowitz72}.

In the spherical geometry the regularized Casimir energy is given by 
\begin{eqnarray*}
E_{Cas}={\lim_{s\rightarrow-\frac{1}{2}}}\mu^{2s+1}\zeta_{H}(s)~~~\mbox{with}~~~ 
\zeta_{H}(s)={\displaystyle \sum_{n=0}^{\infty}{\displaystyle \sum_{l=1}^{\infty}}}(2l+1)(\omega_{n,l}^{2}+m^{2})^{-s}  \label{eq:cassferica}
\end{eqnarray*}
where   $\omega_{n,l}$ are the eingenmodes, and  $\hbar=c=1$ is assumed.
On imposing the boundary conditions \eref{eq:zimp} the eigenfrequencies for the  $TE$ and $TM$ modes are implicitly obtained  \cite{Rosa:2012hj}:
\begin{equation}
\left\{ \begin{array}{c}
\Delta_{\nu}^{TE}(ka):=[\frac{i}{ka}(k a j_{\nu}(ka))'-Z j_{\nu}(ka)]A^{TE}=0\\
\\
\Delta_{\nu}^{TM}(ka):=[-Z\frac{i}{ka}(kaj_{\nu}(ka))'+j_{\nu}(ka)]A^{TM}=0
\end{array}\right.\label{eq:equazconbordimpede}
\end{equation}
so that the  generating function $\Delta_J(k)$ is given by: $\Delta_\nu(k)=\Delta_{\nu}^{TE}\Delta_{\nu}^{TM}$ with $\nu=l+\frac{1}{2}.$
 
In conclusion we can write \cite{Kirsten02}
\begin{equation}
\zeta_{H}(s)=\sum_{\nu=3/2}^{\infty}\frac{\nu}{\pi i}\oint_{\gamma}dk(k^{2}+m^{2})^{-s}\frac{\partial}{\partial k}log[(ka)^{-2\nu}\Delta_{\nu}(ka)].\label{eq:zetmod}
\end{equation}
Shifting the integration contour along the imaginary axis  $k\rightarrow ik$,  we get the following formulae valid in the strip 
  $1/2\mathcal{<R}(s)<1$:
\begin{eqnarray}
\zeta_{H}(s) &=& \frac{\sin(s\pi)}{\pi}\sum_{\nu=3/2}^{\infty}\nu\int_{m}^{\infty}dk(k^{2}-m^{2})^{-s}
\frac{\partial}{\partial k}\log\left[(ka)^{-2\nu}\tilde{\Delta}_{\nu}(ka)\right] \label{eq:zetrass}\\
\nonumber \\
&=&\frac{\sin(s\pi)}{\pi}\sum_{\nu=3/2}^{\infty}\nu\int_{\frac{ma}{\nu}}^{\infty}dy\left[\left(\frac{y\nu}{a}\right)^{2}-m^{2}\right]^{-s}
\frac{\partial}{\partial y}\log\left[(y\nu)^{-2\nu}\tilde{\Delta}_{\nu}(y\nu)\right] \nonumber 
\end{eqnarray}
where  $\tilde{\Delta}_{\nu}=\tilde{\Delta}^{TE}_{\nu}\tilde{\Delta}^{TM}_{\nu}$ is the mode generating function written along the imaginary axis and the $\tilde{\Delta}^{TE(TM)}_{\nu}$ are given by
\begin{eqnarray*}
\tilde{\Delta}_{\nu}^{(TM)}(y)=-Z\left({i y}/{a}\right)\left[\frac{1}{2y}I_{\nu}(y)+\,\dot{I}_{\nu}(y)\right]+I_{\nu}(y)\\
\nonumber\\
\tilde{\Delta}_{\nu}^{(TE)}(y)=\left[\frac{1}{2}-Z\left({i y}/{a}\right)\, y\right]I_{\nu}(y)+y\,\dot{I}_{\nu}(y), \label{eq:newmodeteTM}
\end{eqnarray*}
$I_{\nu}(y)=exp(-i\nu\frac{\pi}{2})J_{\nu}(i y)$ being the modified Bessel functions \cite{Abramowitz72}.

These formulae clarify the role played by the mass $m$. Indeed, the representation \eref{eq:zetmod} is not defined for any  value of $s$ if $m=0$. Thus we will do all the calculations with $m\neq0$ and only in the end we let $m$ go to zero.

Unfortunately  we need to compute $\zeta_{H}(s)$  to the left of the strip  $1/2\mathcal{<R}(s)<1$ 
where it is not defined.
The general technique \cite{Kirsten02} to overcome this problem is to
 add and subtract the asymptotic term
\begin{equation}
h_{as}(y,\nu)=\log\left[(y\nu)^{-2\nu}\tilde{\Delta}_{\nu}(y\nu)\right]_{\nu\rightarrow\infty} \label{eq:termaszet}
\end{equation}
to the integrand so to move the strip of convergence to the left. In this way  an asymptotic zeta function, $\zeta_{as}(s)$, 
is defined. If we are able to compute analytically the asymptotic term and, at the same time, 
 to treat, at least numerically, the remaining one we can find the result.
In particular, let us define 
  $\zeta_{N}(s)$
  as:
\begin{equation}
\zeta_{H}(s)-\zeta_{as}(s)+\zeta_{as}(s)=:\zeta_{N}(s)+\zeta_{as}(s).\label{eq:terminefin}
\end{equation}
We will compute  $\zeta_{as}(s)$ analytically,  while the remaining part, $\zeta_{N}(s)$, must be computed numerically but, because in general it is very small,  we will ignore it.

A careful study of the various terms constituting $\tilde{\Delta}^{TE}_{\nu}~(\tilde{\Delta}^{TM}_{\nu})$ \cite{lucia:2013} showed that a lot of cancellations occur between the  asymptotic contributions of $\tilde{\Delta}^{TE}_{\nu}$ and ~$\tilde{\Delta}^{TM}_{\nu}$. 
Thus, from a computational point of view, it is worth to use $\tilde{\Delta}_{\nu}$ instead of $\tilde{\Delta}^{TE}_{\nu}, \; \tilde{\Delta}^{TM}_{\nu}$ separately.
This simplification allowed us to fix some errors present in \cite{Rosa:2012hj} and to 
point out some problems connected with the emerging divergences (see conclusions).

 \subsection{The Drude model}
 
In this section we compute the Casimir energy for a sphere whose properties  are characterized by the surface impedance  $Z$ of a Drude model \cite{lucia:2013}, i.e.
\begin{equation}
Z_{Drude}(i\frac{y\nu}{a})=\sqrt{\frac{y(y+\sigma_{\nu})}{y(y+\sigma_{\nu})+\delta_{\nu}^{2}}}\label{eq:SIDRUDE}
\end{equation}
where $\omega_{P}$ is the plasma frequency of the material,   $\gamma$ is the relaxation parameter, 
 $\delta_{\nu}=\frac{y_{a}}{\nu}$,  $y_{a}=a\omega_{P}\sqrt{\varepsilon\mu}$,
  $\sigma_{\nu}=\frac{d_{a}}{\nu}$, 
  $d_{a}=a\gamma\sqrt{\varepsilon\mu}$.

To obtain the  asymptotic values of 
  $\tilde{\Delta}_{\nu}$ for  $\nu\rightarrow\infty$ for fixed  $\frac{k}{\nu}$ we make use of the following  uniform asympotic expansion of the Bessel functions \cite{Abramowitz72}:
 \begin{eqnarray*}
I_{\nu}(y\nu)=\frac{1}{\sqrt{2\pi\nu}}\frac{e^{\nu\eta(y)}}{(1+y^{2})^{\frac{1}{4}}}[1+\sum_{k=1}^{\infty}\frac{u_{k}(t)}{\nu^{k}}]\\
\dot{I}_{\nu}(y\nu)=\frac{1}{\sqrt{2\pi\nu}}e^{\nu\eta(y)}\frac{(1+y^{2})^{\frac{1}{4}}}{y}[1+\sum_{k=1}^{\infty}\frac{v_{k}(t)}{\nu^{k}}] \label{eq:espansasin}
\end{eqnarray*}
with  $t=\frac{1}{\sqrt{1+y^{2}}}$,  $\eta=\sqrt{1+y^{2}}+\log(\frac{y}{1+\sqrt{1+y^{2}}})$ and  $v_{k}(t)$ and  $u_{k}(t)$
 are the  Debye's polynomials defined by the following recurrence relation:
 
 \begin{equation}
\left\{ \begin{array}{c}
u_{k+1}(t)=\frac{t^{2}(1-t^{2}}{2})u_{k}^{'}(t)+\frac{1}{8}\int_{0}^{t}dz(1-5z^{2})u_{k}(z)\\
\\
v_{k}(t)=u_{k}(t)-t(1-t^{2})[\frac{1}{2}u_{k-1}(t)+tu_{k-1}^{'}(t)]
\end{array}\right.\;\; k=0,1,2...\label{eq:debpol}
\end{equation}
Inserting this expansion into the expression \eref{eq:zetrass} and developing for 
  $\nu\rightarrow\infty$
 we obtain the asymptotic expression for the $\zeta$-function:
\begin{equation}
\zeta_{as}(s)=\frac{\sin(s\pi)}{\pi}\sum_{l=3/2}^{\infty}\nu\int_{\frac{ma}{\nu}}^{\infty}dy
\left[\left(\frac{y\nu}{a}\right)^{2}-m^{2}\right]^{-s}h_{as}(\nu,y) \label{eq:zetas}
\end{equation}
where $h_{as}(\nu,y)$ can be written as:
\begin{equation}
h_{as}(y,\nu)=\sum_{i=-1}^{n_{max}}\frac{D_{i}(y)}{\nu^{i}}.
\label{eq:cofftot}
\end{equation}
The coefficients $D_{i}(y)$ are given in appendix  up to $n_{max}=5$

\section{Results and conclusions}

All the integrals in \eref{eq:zetas} can be computed using Eq.  \eref{eq:ipergeo}
given in the appendix. In this way, after developing around $m=0$, summing over 
$\nu$ and developing around $s=1/2$ \cite{Rosa:2012hj}, we obtain for the Casimir energy:
\begin{equation}
E_{Cas}=
   \frac{ 1}{a} \Biggl[-0.328-0.504{y_a}^4  -0.441{ d_a}
    { y_a}^4\Biggr]+ E_{div} 
\end{equation}
where
\begin{eqnarray}
&&E_{div} =   \frac{1}{a}\Biggl[ - 0.111\log (a \mu)-\frac{0.055}{   \left(s+\frac{1}{2}\right)}+
  {y_a}^4 \left(0.318 \log
   (a \mu)+\frac{0.035}{s+\frac{
   1}{2}}\right)+ \nonumber\\
 && \;  { d_a}
    { y_a}^4\left(0.132
   \log(a \mu)+\frac{0.330}{s+\frac{1
   }{2}}\right)+ \left(0.248
    { y_a}^4-0.528
    { d_a}  { y_a}^4\right)
   \log (\frac{m}{\mu}) \Biggr]+\nonumber \\
 & &\;  \frac{1}{a}\sum_{j=1}^4 \frac{c_j}{(a m)^j} \label{eq:ecasdiv}
\end{eqnarray}
with the various coefficients $c_j$ explicitly given in the appendix.
We observe that the finite part of the Casimir energy is always negative and that the first corrections due to the material are at least of the  fourth ($y_a^4$) and fifth ($d_a y_a^4$) order respectively. The structure of the divergencies appear, in a sense, more conventional than the one obtained in \cite{Rosa:2012hj}.
Indeed both the  terms $1/(s+1/2)^2$ and  $\log^2{(a\mu)}$ disappeared. This time all the (standard) divergencies are linear and can be eliminated by computing the principal part of the zeta function as usual in the zeta function regularization \cite{Blau:1988lr}.  Very interesting is the the new term appearing in $E_{div}$: $ \frac{1}{a}\sum_{j=1}^4 \frac{c_j}{(a m)^j} $. This term is completely absent in \cite{Rosa:2012hj}, as far we understand it is originated by the fact that the asymptotic expansions we used are non uniform with respect to $y_a (d_a)\in[0,\infty]$.
For this reason to obtain the limit for $y_a\simeq\infty$, which reproduces the case of ideal conducting surface, it is necessary to develop with respect $y_a\simeq\infty$ first, and then  perform the asymptotic expansion with respect to $\nu$. In this way we find:  
\begin{eqnarray}
E_{Cas}=\frac{1}{a} \left\{ 0.084+0.008\log (a)+\frac{0.004}{
   \left(s+\frac{1}{2}\right)}+\right. \\
\left.\frac{1}{ {y_a}}\left[ 0.070+{d_a} \left(-0.038-0.075 \log \left( \frac{a}{m} \right)+\frac{0.001}{
   \left(s+\frac{1}{2}\right)}\right)\right]
\right \} \nonumber
\end{eqnarray}
in agreement with \cite{Leseduarte:1996,Milton:1983}. 

In conclusion in this paper we extended the results of  \cite{Rosa:2012hj}, obtained for the plasma model, to the case of the Drude model.
We developed the calculation up to $n_{max}=5$. The simultaneous treatment of the $TE$ and $TM$ generating functions allowed for a great simplification of the calculations. In this way we could fix some errors present in \cite{Rosa:2012hj}. However, to
better understand the structure of divergencies it would be desirable to have a {\em uniform} asymptotic expansion  over the whole interval $y_a(d_a)\in[0,\infty]$ 
\cite{inprep}.

The possibility of obtaining negative values for the (finite part) of the Casimir energy (attractive force) is confirmed. The approach seems to be very general but the appearance of divergencies of increasing order with respect to the negative power of $m$ deserves further  analysis. The possibility of regularizing the obtained energy remains still an open question and it is currently under investigation.

\section{APPENDIX  }
In the following $\xi=\sqrt{1+y^2}$
 \begin{eqnarray*}
D_{-1}(y) &=&\frac{2}{y} (\xi -1);~~~ 
D_{0}(y)=-\frac{(\xi +y)^2}{\xi ^2 y};~~~ 
D_{1}(y)=\frac{1-y^2}{\xi ^4}-\frac{5 y^3}{4 \xi ^5} \\
D_{2}(y)&=&\frac{{ y_a}^2}{y^3}+\frac{-9 y^5+10 y^3+4 y}{4 \xi^8}+\frac{-9 y^4+14 y^2-2}{4\xi ^7}\\
D_{3}(y)&=&-\frac{3 { d_a}
   { y_a}^2}{2
   y^4}+\frac{-401 y^7+928
   y^5+112 y^3-112 y}{64 \xi
   ^{11}}+\frac{1-25 y^6+70
   y^4-24 y^2}{4 \xi ^{10}} \\
D_{4}(y)&=&\frac{2  { d_a}^2
    { y_a}^2}{y^5}+ 
   \frac{36 y-341
   y^9+1330 y^7-376 y^5-316
   y^3}{16 \xi
   ^{14}}+\\
 & &     { y_a}^4
   \left(\frac{4 y^2+3}{4 \xi 
   y^4}-\frac{1-y^2}{y^5}\right)+\frac{-1363 y^8+5980
   y^6-4292 y^4+512 y^2-8}{64
   \xi ^{13}} \\
 D_{5}(y) &=&  -\frac{5  { d_a}^3  { y_a}^2}{2 y^6}+ { d_a}  { y_a}^4 \left(-\frac{1}{y^2}+\frac{5-6
   y^2}{2 y^6}+\frac{-2 y^4-7 y^2-4}{2 \xi  y^5}\right)+   \\
&&  { y_a}^4 \left(-\frac{3}{8
   y^4}+\frac{8 y^4+13 y^2-7}{8 \xi ^4 y^2}+\frac{-2 y^4-3 y^2-2}{2 \xi ^3
   y^3}\right)+ \\
&&   \frac{-43085 y^{11}+252384 y^9-210320 y^7-56784 y^5+33568 y^3-1312 y}{512 \xi
   ^{17}}+ \\
&&   \frac{-1346 y^{10}+8545 y^8-10660 y^6+3027 y^4-151 y^2+1}{16 \xi ^{16}}.
\end{eqnarray*}
The relevant integral formula:
\begin{eqnarray}
\int_{\frac{m a}{\nu}}^\infty{\frac{y^b}{\left(1+y^2\right)^c}\left( \frac{\nu^2y^2}{a^2}-m^2\right)^{-s}}dy &=&
 \frac{  \left(\frac{\nu ^2}{a^2}\right)^{c-\frac{b+1}{2}}\Gamma (1-s)}{2\left(m^2\right)^{s+c-\frac{b+1}{2}} } \Gamma\left(c+s-\frac{b+1}{2} \right) \label{eq:ipergeo}
\\
&&
   {_2{F}}_1\left(c, c+s-\frac{b+1}{2},c-\frac{1-b}{2};-\frac{\nu ^2}{a^2m^2}\right)  \nonumber 
\end{eqnarray}
where $_2F_1$ is the hypergeometric function \cite{Abramowitz72}.
The $c_j$ coefficients in Eq. \eref{eq:ecasdiv}
 \begin{eqnarray*}
 c_1 &=&   0.818  { d_a}
    { y_a}^4+0.234
    { y_a}^4+0.115
    { y_a}^2\\
    c_2& =&  0.080
    { y_a}^4-0.073
    { d_a}  { y_a}^2 \\
   c_3&=&0.057
    { d_a}^2  { y_a}^2-0.125
    { d_a}
    { y_a}^4-0.0286
    { y_a}^4 \\
 c_4 &=&{0.0486  { d_a}{ y_a}^4-0.0486
    { d_a}^3  { y_a}^2} 
 \end{eqnarray*}
 
%
\section{References}
\bibliographystyle{unsrt}

\bibliography{bibliografia}
\nocite{*}

\end{document}